\documentstyle[11pt,paspconf,epsfig]{article}

\begin{document}

\title{On the Particle Heating and Acceleration 
in Black Hole Accretion Systems}

\author{H. Li\altaffilmark{1}, S. A. Colgate\altaffilmark{1},
 M. Kusunose\altaffilmark{2} and R.V.E. Lovelace\altaffilmark{3}}
\altaffiltext{1}{Theoretical Astrophysics, T-6, MS B288, Los Alamos
National Laboratory, Los Alamos, NM 87545, hli@lanl.gov, colgate@lanl.gov}
\altaffiltext{2}{Department of Physics, School of Science,
Kwansei Gakuin University, Nishinomiya 662-8501, Japan,
 kusunose@kwansei.ac.jp}
\altaffiltext{3}{Department of Astronomy, Cornell University,
Ithaca, NY 14853, rvl1@cornell.edu}

\begin{abstract}
The lack of our knowledge on how angular momentum is transported
in accretion disks around black holes has prevented us from fully
understanding their high energy emissions. We briefly highlight
some theoretical models, emphasizing the energy flow and electron
energization processes. More questions and uncertainties
are raised from a plasma physics point of view.  

\end{abstract}


\section{Introduction}

\begin{figure}
\epsfig{file=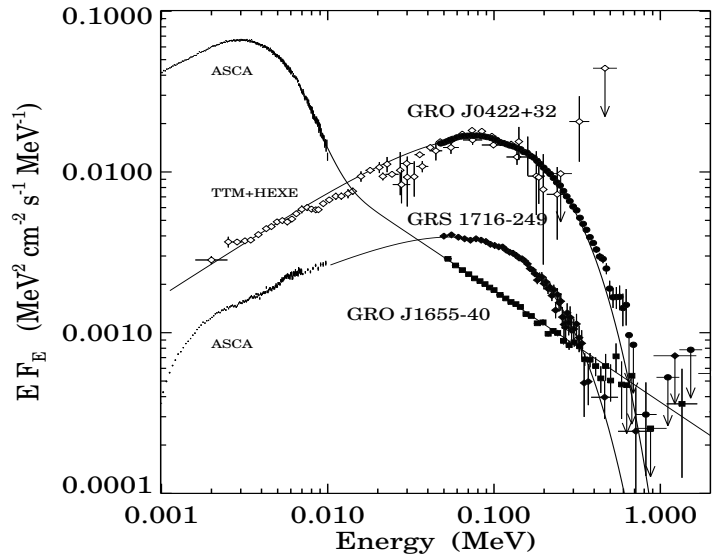,width=10cm,height=8cm}
\caption{Broad band $\nu F_{\nu}$ spectra for 
three GBHCs. They indicate the low-hard and high-soft states
commonly seen from these objects. Note the excess of emissions
above 500 keV in both states. Taken from Grove et al. (1998). } 
\label{fig-1}
\end{figure}

Figure \ref{fig-1} shows three (roughly) contemporaneous broad band high 
energy emission spectra from three galactic black hole candidates 
(GBHCs; Grove et al. 1998). Although it is conventional to interpret the
soft black-body-like component below $\sim 10$ keV as coming from
an optically thick Shakura-Sunyaev (SS) disk, the origin of the
hard X-ray continuum (and its extension into soft X-rays during the
low-hard state) is a constant source of debate. Extracting a physically
sensible model through a maze of high quality spectral and timing data 
on these systems remains a great challenge. 

Recently, there seems to be a renewed interest in understanding particle
heating/acceleration in accretion disks. We attribute this to the
observations of: possible $> 0.5$ MeV emissions from Cyg X-1 and GRO J0422;
the powerlaw component of GRO J1655  extending
to at least 800 keV without a cutoff (Tomsick et al. 1998); and
relativistic radio jets from sources like GRO J1655 and GRS 1915.
Furthermore, the clearly laid-out physical requirements of 
ADAF models (which have enjoyed much success, see Narayan et al. 1998 for
a review) also prompted further discussions on particle heating.

In this review we will mostly discuss a few models
for the so-called
low-hard state where the spectrum ($\nu F_\nu$)
is peaking around 100-200 keV. We 
apologize for not able to cover all the models
(see Liang 1998 for a recent extensive review).
 The powerlaw tail that seems to
extend beyond $500$ keV during the soft-high state also
begs explanation, though the total energy contained in this tail
is perhaps $< 10\%$ of the total emission, so we will place less
emphasis on them. We will focus on the
electron energization processes of these
theoretical models. We will not discuss
any detailed spectral and temporal analyses (see other articles
in this volume). Even so, we quickly realized
that writing on this topic is a very difficult task because we 
find many questions and confusions with no clear and definite answers.

\section{Some Models for the Origin of Hard X-rays and Gamma-rays} 

In all the models discussed here, the physics of angular momentum
transport (or ``$\alpha$'' viscosity) during accretion is not 
well understood. 
As a direct consequence, unfortunately, modeling 
energy dissipation in accretion disks has many {\em ad hoc} elements.
Quite generally, the matter (surface density $\Sigma$) in accretion 
disk is evolved as (taken from Papaloizou \& Lin 1995)  
\begin{equation}
{\partial \Sigma \over \partial t} -
\frac{1}{r} \frac{\partial}{\partial r}\left[F_1 + F_2 + F_3\right]
-S_{\Sigma} = 0
\end{equation}
\noindent where $F_1 \propto \partial 
(\langle \nu \rangle \Sigma r^{1/2})/\partial r$ is the 
local viscous transport with viscosity $\langle \nu \rangle$
(i.e., the standard $\alpha-$disk viscosity or from MHD turbulence
by Balbus \& Hawley 1991, 1998); $F_2 \propto S_{\Sigma}J$ is the advective
loss with $J$ being the angular momentum carried by the source/sink 
($S_{\Sigma}$) material (i.e., magnetic flux and/or winds, 
Blandford \& Payne 1982); $F_3 \propto \Lambda$ is the external 
perturbation (i.e., tidal interactions).

Three models (or their variants) are usually employed for explaining
the high energy emissions, namely, the SS model, the SLE model
(Shapiro et al. 1976),
and the ADAF model. All of them use the local viscous transport
prescription (the $F_1$ term) and the energy is also dissipated
locally at the disk. In SS model disk is optically thick and
geometrically thin, and the plasma is also highly collisional.
The heat deposited from transporting angular momentum is
successfully radiated away so that disk remains thin ($H \ll R$).
In SLE and ADAF models, however, an inner, hot ($T_e \sim 100$ keV),
optically thin ($\tau \leq 1$) and two-temperature ($T_i \gg T_e$)
region is postulated. This region is then cooled via various 
radiation processes, such as thermal Compton scattering and Synchrotron.

The arguments for the existence of this hot, optically thin region
might be summarized as follows: if local viscous energy dissipation
{\em only} heats protons, and if there is only Coulomb coupling
between electrons and protons, then when the energy input rate is high 
enough, the system will become unstable if the cooling via radiation
is not quick enough, so the plasma has to expand and become
optically thin. Here, we want to emphasize that the accreting
plasma, during this transition from an optically thick, thin disk 
to an optically thin, quasi-spherical state, has also changed
from {\em highly collisional} to essentially {\em collisionless}.
This brings up several immediate questions which are related
to the above ``if''.  

\section{Open Questions}

\subsection{Will local viscous energy dissipation 
only heat protons?}

Bisnovatyi-Kogan \& Lovelace (1997) first discussed
this issue and argued that dissipation in such a
magnetized collisionless plasma predominantly heats
the electrons owing to reconnection of the
random magnetic field. On the other hand,
Quataert (1998) and Gruzinov (1998) have argued that
conditions for ADAF could be true in the high
$\beta = P_{\rm plasma}/P_{\rm magnetic} \geq 5$ limit
by calculating the linear damping rates of
short wavelength modes in a hot (but nonrelativistic) plasma.
in an (implicit) almost uniform magnetic field.
Note that even though MHD turbulence phenomenology 
was used in both papers, the damping rates are 
valid in the linear regime for plasma waves only 
(see below for further discussion).
But these calculations perhaps are not answering the question 
of how to form the optically thin region in the first place
because they are damping rates in the {\em collisionless} limit.
Instead, one perhaps might first evaluate the energy dissipation 
processes (with an understanding of $\alpha$ viscosity) in the 
{\em collisional} limit which is the physical state initially. 
These collisions ensure thermal electron and proton distributions 
and efficient energy exchange between them, especially
at the so-called transition radius in ADAF ($10^3-10^4 r_s$). 

If one uses Balbus-Hawley instability (see also 
Velikov 1959 and Chandrasekhar 1981) as the origin of the viscosity
in the disk, then the gravitational energy is mostly released
in large scale (longest wavelength of the magnetic field
changes) and this energy will amplify the field first (instead
of going into heating the particles). Once the nonlinear
saturation is reached (say with magnetic energy density being
$10\%$ of the kinetic energy density of the shear flow), we are
actually faced with two possibilities, namely, whether
the magnetic fields will be expelled (or escape) from the disk,
or they will have to dissipate locally in the disk.
Bisnovatyi-Kogan \& Lovelace (1997) argued for the second
possibility (but see Blackman 1998).
Since we know that both the fluid and magnetic Reynolds numbers
are exceedingly large in these flows, any ``classical'' 
viscous and ohmic dissipations
will happen on timescales longer than the age of the universe, thus
efficient magnetic reconnection has been 
sought as the primary candidate 
for energy dissipation in the disk. They further argued that
current-driven instabilities in this turbulent plasma will
give rise to large local $E_{\parallel}$ which mostly accelerate
electrons. Thus, up to half of the magnetic energy input goes
directly to electrons and is subsequently radiated away, and
the disk will always stay thin and optically thick. The
uncertainties in these arguments are nevertheless quite 
large since we don't fully understand MHD turbulence, let
alone its dissipation via kinetic effects. For example, it
is unclear whether such reconnection sites are populated
throughout the plasma so that most fluid elements encounter
such regions. There has been
some detailed numerical simulations with magnetic Reynolds number
up to 1000 (Ambrosiano et al. 1988)
in which test particles are observed to get accelerated by
the induced small scale electric fields associated with
reconnection sites in turbulent MHD flows.
If indeed the magnetic energy dissipation is through
accelerating particles by the induced electric fields
(this is a big if), since electrons are the current carriers, 
it is hard to imagine that protons receive most of the energy.

\subsection{Is there any collective process that could
ensure efficient energy exchange between protons and electrons 
besides Coulomb?}

Putting aside the uncertainties discussed above, if there is 
indeed an optically thin, hot, two-temperature plasma region, 
a pertinent question is how much energy electrons can get.
This question is, unfortunately, ill-fated again because we do
not know how to formulate the problem. Another way to look at it
is how to identify the free energy, since most
plasma instabilities require a good knowledge of the free energy
as determined by the system
configuration. For example, is there a relative drift between
protons and electrons and can fast electrons be regarded as
a beam to an Maxwellian proton core distribution? Is there
temperature anisotropy parallel and perpendicular to background
magnetic fields, etc.? Begelman \& Chiueh (1988) have studied some
plasma instabilities in detail and found plausible ways of
transferring energy from ions to electrons, under the
conditions that a substantial level of MHD turbulence will
give a large enough proton density gradient (or curvature drifts)
so that proton drift velocity can be large enough to drive
certain modes unstable. The fluctuating electric field parallel
to the magnetic field will then accelerate electrons. The applicability
of this instability is again hampered by our lack of knowledge
of the presumed MHD turbulence. Narayan \& Yi (1995) argued
that this mechanism does not work well in ADAF.
 
A conceptual difficulty is that the typical modes excited by
protons (having most of the energy) are below the proton gyrofrequency 
$\Omega_{ci}$. This makes resonance with the electrons difficult.
But a possible avenue is to have protons excite (almost) perpendicular
modes (i.e., high $k_\perp$ and very small $k_\parallel$). Then the
resonant conditions for electrons to resonate with these waves are
easier to satisfy. More work is needed to explore these possibilities. 

\subsection{Could accretion disk have a magnetically
dominated, hot corona like our Sun?}

The formation of a ``structured corona'' was first proposed  by
Galeev et al. (1979). In this model a radial quadrupole field is 
wound up by differential rotation into
an enhanced toroidal field. Then the helicity of the presumed convective
``turbulence'' converts a fraction  of the toroidal flux back 
into poloidal field and hence produces an exponentiating dynamo that 
saturates by back reaction. This is the classical
$\alpha - \Omega$  dynamo although not identified as same in the paper.
Furthermore the saturation or back reaction limit of this disk
dynamo is assumed to be the random loops of flux characteristic 
of the solar surface. 

One important step in the above model is the requirement
of vertical (thermal) convection in the $\{R,z\}$ plane.
The convective motion may be driven by heat released at or near 
the mid-plane. Lin et al. (1993) have shown that under
specific conditions of opacity and equation of state that convective
instability should occur both linearly and nonlinearly, thus  
leading to large amplitude cells. However, the convective cells  
are highly constrained radially. The problems of restrictive
initial conditions and the restrictive cell geometry 
leads one to conclude that this is not the universal
mechanism needed to explain accretion disks.  
Colgate \& Petschek (1986) showed
that to drive convective cells whose displacement radially is of 
the order of the disk height $h$,  (unrealistic)  
efficiency of the energy flow (a Carnot cycle of $\sim 100\%$ 
efficiency) is necessary to drive these convective eddies, and
the cells created are also highly restrictive, tall but
narrow radially (i.e., similar as Lin et al.).
Thus the existence of strong convective turbulence is doubtful. 
The result of this lack of convective turbulence with  rising
plumes is to negate the origin of the helicity invoked in the 
structured corona model.

\section{Electron Energization}
 
Besides the possible role of magnetic reconnection in accelerating
electrons which is observed in the solar corona (Tsuneta 1996),
there are more standard processes which involve
wave-particle interactions (see Kuijpers and Melrose 1996). 
Shock acceleration is not considered here.
We give a quick review of the electron energization by plasma
waves and turbulence. 

\subsection{Particle heating/acceleration -- linear and quasilinear
theory}

Linear Vlasov equation is usually used to describe the collisionless
plasma, which is a good approximation of astrophysical plasmas.
Linearization of the Vlasov equation yields various dispersion
relations  $\omega = \omega({\bf k})$ which describe
how the system will respond to {\em small} electrostatic and electromagnetic
perturbations. Since the field energy of low frequency
fluctuations (i.e., $\omega < \Omega_{\rm cp}$) is predominantly
magnetic, particles generally experience strong pitch-angle scattering before
they can be energized. Of  fundamental importance is the
wave-particle resonance, that is, given an electromagnetic fluctuation
of frequency $\omega$ and wavevector ${\bf k}$, a charged particle ($q,
m$) is considered to be in resonance with this fluctuation when

$$ \omega - k_{\parallel} v_{\parallel} - \ell \Omega_0 / \gamma =
0, ~~~~~~~~~~\ell = 0, \pm 1, \pm 2,~ \dots \eqno(1) $$

\noindent where the nonrelativistic gyrofrequency $\Omega_0 = |q|B_0/mc$, 
and $v_{\parallel}$ and $\gamma$ are the particle's parallel
velocity and Lorentz factor, respectively. When the harmonic number 
$\ell = 0$, the resonance is referred to as the Landau or Cherenkov 
resonance, and
implies that the particle speed along the magnetic field matches the 
speed of the parallel wave electric or magnetic field. If $|\ell| > 0$, 
the process is called gyroresonance, and there is a matching between the wave 
transverse electric field and the cyclotron motion of the particle. 
The sign of $\ell$ 
depends upon the transverse polarization of the wave and the sign 
of $q$: if the transverse wave electric field and the particle rotate 
in the same
 sense about $B_0$ in the plasma frame, then $\ell$ is positive. 
In most settings, only $\ell = \pm 1$ is of importance.

The key quantities are the plasma $\beta=n(T_i+T_e)/(B^2/8\pi)$
factor and the temperature ratio $T_e/T_i$. Furthermore, we have:

$\bullet$ Linear theory. The linear theory of plasma waves and
instabilities is often reduced to a linear dispersion equation
with a complex $\omega$, whose imaginary part gives the growth
or damping of certain modes. Recent studies by Gruzinov (1998)
and Quataert (1998) belong to this case. The usual candidates
for wave-particle resonances are: for $\ell=0$, the transit 
time damping (TTD) for particle with oblique fast magnetosonic waves and 
Landau damping (LD) with kinetic Alfven waves; for $|\ell| \geq 1$,
gyroresonances between the proton/Alfven wave and the electron/whistler wave. 

$\bullet$ Quasilinear theory. 
A detailed physical understanding of pitch-angle scattering and
stochastic acceleration is beautifully presented in Karimabadi et
al. (1992), using nonlinear orbit theory with the
Hamiltonian formalism. In the presence of a continuum of
plasma waves, the number of
resonances between the particle and waves is greatly increased to a
point that the trapping width associated with one particular resonance
can overlap with neighboring resonances, thus allowing particles
``jump'' from one resonance to another. As particles sample different
resonances, they gain energy in a ``ladder-climbing'' fashion. Hence
the description stochastic acceleration. This approach has been
adopted in several studies on electron acceleration by fast-mode
waves and whistler waves in accretion disk (Li et al. 1996, Li
\& Miller 1997). We typically find that the electron distribution is
hybrid with a nonthermal tail, which is responsible for the
production of $> 500$ keV emissions in several GBHCs.
We have built a computer code which solves 3 coupled, time-dependent 
kinetic equations for particles, photons and waves, respectively.
Namely,
\begin{equation}
\frac{\partial N_e}{\partial t} 
= -\frac{\partial}{\partial E} \left\{
\left[\Big\langle \frac{dE}{dt}\Big\rangle
+ \left(\frac{dE}{dt}\right)_{\rm loss} \right] N_e \right\} 
+ \frac{1}{2} \, \frac{\partial^2}{\partial E^2}
\left[(D + D_c) N_e \right] 
\end{equation}
\begin{equation}
\frac{\partial W_{\rm T}}{\partial t} = \frac{\partial}{\partial k}
\left[
      k^2 D \, \frac{\partial}{\partial k} \, 
      \left( k^{-2} W_{\rm T} \right) 
\right]
-\gamma W_{\rm T} +
Q_{\rm W}\delta(k-k_0) 
\end{equation}
\begin{equation}
\frac{\partial n_{\rm ph}(\varepsilon)}{\partial t} 
= - n_{\rm ph}(\varepsilon) \, \int dE \, N_e(E) \, 
R(\varepsilon, E) + 
\end{equation}
$$
\int\int d\varepsilon^{\prime} \, dE \, 
P(\varepsilon;\varepsilon^{\prime},E) \,
n_{\rm ph}(\varepsilon^{\prime}) N_e(E) 
 \mbox{} + \dot{n}_{\rm ext}(\varepsilon) 
+ \dot{n}_{\rm emis}(\varepsilon) 
- \dot{n}_{\rm abs}(\varepsilon) 
- \frac{n_{\rm ph}(\epsilon)}{t_{\rm esc}} \, .
$$

The particle distribution can be arbitrary. This allows us
to determine from all the interactions whether the distribution
is thermal or nonthermal. Pair production is not included so far.
The Coulomb terms are also implemented for arbitrary particle
distributions. Accurate Compton scattering is treated as a
scattering matrix (Coppi 1992) with the full cross section.
The Cyclo-Syn. process is calculated according to Robinson \& Melrose
(1984) which enters both as a cooling and heating term
(Ghisellini et al. 1988). Syn-self absorption is also included.
The radiation part of the kinetic code is tested against 
Monte Carlo simulations (Kusunose, Li, \& Coppi 1998) and
is found to be very good for $\tau \leq 3$ and for
both thermal and nonthermal electron distributions.

\begin{figure}[htbp]
\epsfig{file=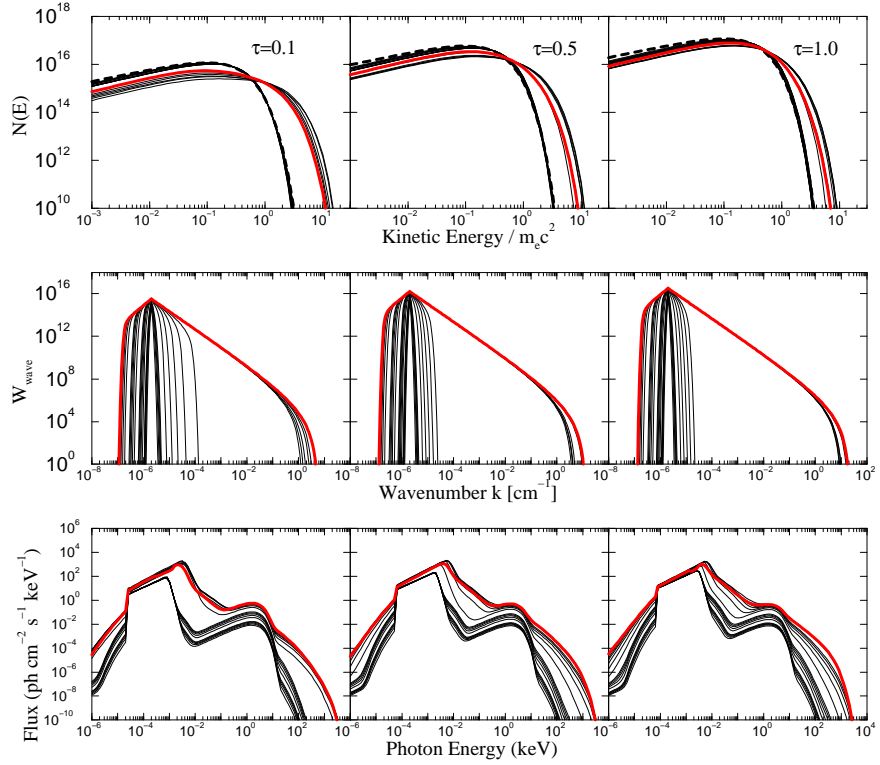,width=10cm,height=12cm,angle=-90}
\caption{Time evolution (from $t=0-10 R/c$) of particle
distribution (top), MHD wave spectral density (middle) and
photon flux (bottom). The initial particle $T_e = 100$ keV
and $\tau = 0.1, 0.5, 1$, respectively. The soft photons
are injected with $T_s = 1$ keV and the compactness $\ell_s = 10$.
The size is $\sim 30 r_g$ and $M= 7 M_{\odot}$. The final
distributions are indicated by thick curves and dashed lines
are initial distributions. Deviations from Maxwellian
are obtained as MHD waves cascade to higher $k$, accelerating
electrons out of the thermal bath. Cyclo-Syn. (with self-absorption)
and Compton are the radiation processes considered here.}
\label{fig-2}
\end{figure}

Figure \ref{fig-2} shows the time evolution of particles (upper panels),
waves (middle) and photons (lower panels) from the start of continuous
wave injection until the steady state is reached.
There are 20 curves in each plot which are from $t=0 - 10 \tau_{dyn}$,
where $\tau_{dyn} = R / c \sim 1.5\times 10^{-3}$ sec.
These runs are made with application to optically thin environment
in mind. 
The plasma density $n$ is varied from $\tau = 0.1 - 1$.
At early times, the particle distribution softens first
as shown in upper panels, due to that waves have not fully cascaded
(i.e., small $\langle k \rangle$ as shown in middle panel), and losses
dominate at high energies. As waves cascade over the inertial range,
$\langle k \rangle$ quickly grows to a level that
acceleration overcomes all losses, electrons are then energized out of
the thermal background and the nonthermal hard tail forms.
The photon spectra  indicates
that  gamma-rays can be produced when $\tau < 0.5$.
Furthermore, note that the nonthermal tails start to develop
at $E/m_e c^2 \sim 0.13$ (corresponding to $v_{\rm A}/c = 0.46$),
this nicely confirms the fact that only particles with
$v > v_{\rm A}$ can be accelerated.

\subsection{MHD turbulence, are they an ensemble
of waves?}

The above described calculations, both linear and quasilinear,
can be broadly regarded as ``dissipation'' in a general MHD 
turbulence theory. Finding a dynamical model that might
adequately describe the evolution of magnetic fluctuations
(such as equation (3) above)
is at best phenomenological. In the dissipation range, the
physics of the couplings that connect fluid and kinetic
scales is not understood at all.
 
A critical assumption that is employed in all the kinetic calculations
is that the magnetic fluctuations that cascade from large scales
to small scales could be regarded as an ensemble of kinetic waves
with a well-defined dispersion relation to describe them.
This view is by no means proven, though it allows us to 
get an estimate of the particle heating rate since the kinetic
theory is significantly more advanced (see an application of
such an approach to the interplanetary magnetic field 
dissipation range, Leamon et al. 1998). 
On the other hand, the dynamics of MHD turbulence has been
studied using statistical theories and simulations
(e.g., Kraichnan \& Montgomery 1980; Shebalin et al. 1983;
Matthaeus \& Lamkin 1986),
and has never been convincingly presented
or developed within a normal-mode, perturbation-type of framework.
A further complication is that most (MHD) turbulence theory
is based on the incompressible fluid model, how it will ``carry-over''
to compressible astrophysical flow is still an open question.

\subsection{MHD turbulence Truncation}

Recently,  the assumption of a cascade to smaller scales of
MHD turbulence is criticized in dynamo theory. It has been argued
that both the more rapid folding of magnetic flux as well as the 
smaller energy density at small scale ensures rapid saturation or 
back reaction by the field stress,  immobilizing
the small scale fluid motions expected from the Kolmogorov spectrum.  This
will truncate the turbulent spectrum at the back reaction scale, initially
the smallest and progressively reaching the largest. Since the energy
input to the turbulence is assumed to be primarily at the largest scale,
this leaves one with negligible power at the small scale 
(Cattaneo \& Vainshtein 1991; Kulsrud \& Anderson 1992;  
Gruzinov \& Diamond 1994; Cattaneo 1994). The remaining largest scale is
that of the disk itself.

In general all particle instabilities presumably leading to particle
heating require large local gradients in some aspect of their phase space,
i.e. temperature, density, and velocities, etc. Furthermore, all the free
gravitational energy must flow through these gradients.  This 
requirement, however, will not be met if the small scale turbulent 
motions are strongly damped by the back reaction of the field itself.  
We therefore look for a solution to this paradox in large scale magnetic 
structures.

\section{A Sketch View}

Here, we outline some plausible physical pictures
about what might be happening in an accretion flow. Most these
are ideas that have not been thoroughly investigated.
It is also clear that there are obvious gaps which need to be filled 
with rigorous calculations.

\subsection{Hydrodynamic transport and high-soft state}

Many investigations have sought a linear
instability deriving energy from the Keplerian flow to produce a growing 
mode leading, in the non-linear limit, to turbulence.  The Papaloizou
\& Pringle instability (1984) seems to be the most studied instability
but its relevance to Keplerian accretion disks has been questioned
(Balbus \& Hawley 1998).

Recently we have identified a linear instability in Keplerian
disk leading to Rossby 
waves and presumably Rossby vortices in the nonlinear limit
(Lovelace et al. 1998). This instability grows most effectively 
from a large radial gradient in entropy.  It has the advantage that 
the nonlinear limit consists of co-planar, co-rotating vortices 
(Nelson et al. in preparation) that  
require only a radial gradient, not vertical gradient
of entropy.  The radial gradient, we believe, is astrophysically reasonable
because all disks are presumably fed by matter at some outer radius by,
for example, Roche-lobe overflow in low-mass X-ray binaries. 
If there is no angular momentum removal mechanism, the matter will
accumulate until it builds up enough to trap heat, and
variations in entropy would then render the onset of the 
above instability.
We do not, however, expect this instability to lead to
turbulence in the usual sense of convective turbulence.  
An ensemble of co-planar vortices does not lead to
significant vertical flow as compared to the usual picture of
convective turbulence where buoyant plumes would convect heat 
released at the the mid-plain to the disk surface.

\begin{figure}[htbp]
\epsfig{file=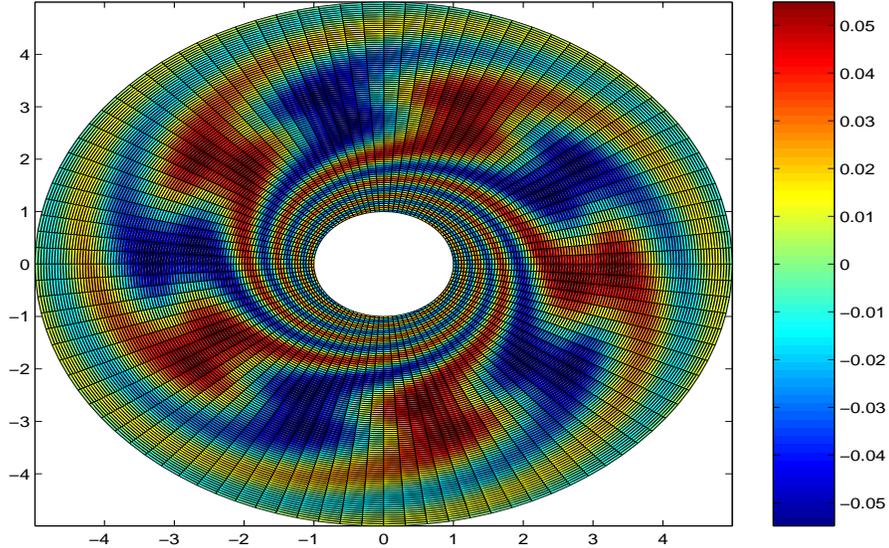,width=12cm,height=7.5cm}
\caption{A linearly  unstable mode in a 2D Keplerian disk with 
an entropy bump initially located at $r=3$. Shown is the amplitude 
($\pm$ means moving out/in) of the radial velocity $v_r$ which is 
zero initially. This is an $m=5$ mode. The unstable mode
is ``trapped'' at the entropy bump. Since the flow is nonbarotropic,
additional vorticity ($\nabla\times {\bf v}$) is also produced
around $r=3$. }
\label{fig-3}
\end{figure}

We expect the angular momentum transport is done via
nonlinear interactions of these vortices with the background
flow, but this has to be addressed by extensive hydro simulations.
The heat flow derived from the ``viscosity'' of the ensemble of
Rossby vortices must be removed by radiation flow.  We expect this not
to be a problem because the
radiation thickness of the disk, $\tau$, is small enough
such that the effective diffusion velocity, $v_{\rm diff} 
\simeq c/(3 \tau) >> v_{\phi}$.
Under these conditions, 
the disk solution will be essentially the same as the SS disk.
Thus, this picture might be applied to the high-soft state
of GBHCs. Relatively speaking, magnetic fields do not play
a major role during this state but some nonthermal processes
(such as a weak magnetic outflow)
might be responsible for the powerlaw component from $20$ keV - 1 MeV.

\subsection{Role of large scale magnetic fields and low-hard state}

As pointed out by Blandford \& Payne (1982), large scale
magnetic fields can also be very effective in removing
the disk angular momentum. These large scale magnetic 
outflows could be a hydromagnetic wind (Blandford \& Payne 1982),
or it might be a nearly force-free helix (Poynting flux)
with very little matter as discussed by 
Lovelace et al. (1987, 1997).

The accreting plasma from, say a companion star, is likely
to be magnetized. In the advection of this flux with the mass flow, 
there will necessarily be a convergence and strengthening of the field. 
In the region where an $\alpha$-viscosity prevails and the field acts 
as a passive marker of the flow, there will be both advection and 
diffusion.  The diffusion radially outwards
depends upon probably the same diffusion coefficient  which allows the
diffusion of angular momentum. Hence there will be a unique
relationship between advection inwards and diffusion outwards leading to the
relationship, $B_{z} \propto r^{-3/2}$ (see also Bisnovatyi-Kogan
\& Lovelace 1997 in which they argued $B_r \propto r^{-2}$).  

If the initial field strength advected with the mass flow
at the outer disk radius is large enough, then the field energy 
density could become comparable to the Keplerian
stress at a certain radius.
Bisnovatyi-Kogan \& Lovelace (1997) argued that magnetic flux then
has to be destroyed at the disk via reconnection. 
Alternatively, instead of destroying the flux, magnetic fields
(presumably tied to the companion star) could be twisted such
as they will remove the angular momentum of the flow and take
away the released gravitational energy. 
So the energy dissipation
(into radiation) might not be at the disk at all.
The reconnection dissipation of the current supporting the torsion 
of the magnetic field
will perhaps lead to the non-thermal emission of GBHCs.
In fact, there is ample evidence in AGNs that perhaps most of
the energy release is in the outflow/jet. Such a picture
could also apply to GBHCs with the hard X-ray to gamma-ray
emissions produced via nonthermal processes (such as Syn. or SSC) 
in the magnetized outflow away from the disk. 

If the initial field strength advected with the mass flow
at the outer disk radius is weaker, then the amplified
magnetic field (by $r^{-3/2}$) may never be greater than
the Keplerian stress, thus is not the dominant channel of
angular momentum transport. However, there will always  be  
nonthermal energy release in the twisted magnetic field,
which could be the powerlaw tail during the high-soft state.

\acknowledgments

HL wish to thank the meeting organizers for their
kind invitation and financial support.
Part of the work on quasilinear wave-particle interactions
was done together with J. Miller. We thank J. Finn and E. Liang
for many useful discussions. HL gratefully acknowledges the
support of an Oppenheimer Fellowship at LANL.

\end{document}